\documentclass[reprint, a4paper, superscriptaddress, amsmath, amssymb, aps, pra, tightenlines, twocolumn]{revtex4} 
\usepackage{color}
\usepackage{graphics}
\usepackage{epsfig}
\usepackage{sansmathaccent}
\begin{document}
	
\title{Coexistence of long-range magnetic order and dynamical magnetism in the V-based Kagome metals: A combined thermodynamic and $\mu$SR study}
	
\author{Sheetal Devi}
\email{sdevi250495@gmail.com}
\altaffiliation{Current Address: Graduate School of Engineering, Kyushu Institute of Technology, Kitakyushu 804-8550, Japan\\}
\affiliation{J\"ulich Centre for Neutron Science (JCNS) at the Heinz Maier-Leibnitz Zentrum (MLZ), Forschungszentrum J\"ulich, Lichtenbergstrasse 1, D-85747 Garching, Germany}
	
\author{Yishui Zhou}
\affiliation{J\"ulich Centre for Neutron Science (JCNS) at the Heinz Maier-Leibnitz Zentrum (MLZ), Forschungszentrum J\"ulich, Lichtenbergstrasse 1, D-85747 Garching, Germany}
\affiliation{Technical University of Munich (TUM), TUM School of Natural Sciences, Physics Department, D-85747 Garching, Germany}

\author{Thomas J. Hicken}
\affiliation{PSI Center for Neutron and Muon Sciences  CNM, 5232 Villigen PSI, Switzerland}

\author{Zurab Guguchia}
\affiliation{PSI Center for Neutron and Muon Sciences  CNM, 5232 Villigen PSI, Switzerland}
	
\author{Hubertus Luetkens}
\affiliation{PSI Center for Neutron and Muon Sciences  CNM, 5232 Villigen PSI, Switzerland}
	
\author{Min-Kai Lee}
\affiliation{Department of Physics, National Cheng Kung University, Tainan 70101, Taiwan}

\author{Lieh-Jeng Chang}
\affiliation{Department of Physics, National Cheng Kung University, Tainan 70101, Taiwan}
	
\author{Yixi Su}
\email{y.su@fz-juelich.de}
\affiliation{J\"ulich Centre for Neutron Science (JCNS) at the Heinz Maier-Leibnitz Zentrum (MLZ), Forschungszentrum J\"ulich, Lichtenbergstrasse 1, D-85747 Garching, Germany}

\begin{abstract}
	V-based Kagome metals exhibit a unique lattice geometry that can give rise to exotic electronic and magnetic phenomena, making them an ideal platform to study the interplay of topology and magnetism. We present a combined thermodynamic and muon spin relaxation ($\mu$SR) investigation of single-crystal \textit{R}V$_{6}$Sn$_{6}$ (\textit{R} = Tb, Dy, Ho, Er) compounds, focusing on their low-temperature magnetic behavior. Heat capacity and $\mu$SR measurements reveal distinct magnetic phase transitions below 4 K, confirming the emergence of long-range magnetic order in all compounds studied. The $\mu$SR results further indicate persistent spin fluctuations within the magnetically ordered state down to 50 mK, reflected in reduced ordered moments obtained from hyperfine analysis of the heat capacity measurements. These findings uncover the coexistence of static and dynamic magnetism in V-based Kagome metals and emphasizing the key role of 4$f$-electron anisotropy in shaping their magnetic ground states. Compared with the Mn-based \textit{R}Mn$_{6}$Sn$_{6}$ analogs, our results highlight the unique magnetism arising from the decoupled rare-earth sublattice and its interplay with the nonmagnetic V Kagome network.
\end{abstract}

\maketitle
	
\section{Introduction}
Metals featuring Kagome lattices have recently attracted considerable attention due to the emergence of distinctive electronic properties, such as saddle points, Dirac cones, and geometrically driven flat bands within the band structures \cite{peng2021realizing,wang2020experimental,hu2023optical,lee2022anisotropic,yin2020quantum}. Many of these materials are considered potential candidates for exhibiting correlation-driven topological magnetism. Particularly within the subset of \textit{R}M$_{6}$X$_{6}$ materials, which feature a Kagome layer of transition metal ions (\textit{M}), they provide significant potential for studying the pristine Kagome lattice under idealized conditions. These materials manifest a complex interplay between the nontrivial effects of the topological band structure arising from the underlying sublattice and the magnetism originating from potentially large 4\textit{f} moments of rare-earth (\textit{R}) sites and 3\textit{d} moments of the transition-metal sites.
	
In the context of magnetic Kagome systems, specifically with \textit{M} = Mn, recent investigations have revealed materials where the magnetic structure of the Mn-sublattice can be flexibly modulated by varying the \textit{R}-site with different 4f moments and ion anisotropy. For instance, ferrimagnetic structures have been observed for \textit{R} ranging from Gd to Ho, while antiferromagnetic structures emerge for \textit{R} = Er, Tm, and Lu \cite{ma2021rare,wang2022magnetotransport,pokharel2021electronic}. Of particular interest is TbMn$_{6}$Sn$_{6}$, which exhibits Chern-gapped Dirac fermions, contributing to anomalous electronic and thermal transport effects \cite{yin2020quantum,xu2022topological}.
	
In contrast to the robustly magnetically ordered \textit{R}Mn$_{6}$Sn$_{6}$ compounds, the \textit{R}V$_{6}$Sn$_{6}$ series, where a nonmagnetic vanadium site substitutes for manganese at the intermetallic \textit{M} sites, may similarly engender significant physics while facilitating a clear separation between the magnetic and electronic subsystems. Magnetic ordering transitions in \textit{R}V$_{6}$Sn$_{6}$ occur at relatively low temperatures, with ordering temperatures below 6 K for \textit{R} = Gd to Er and no observed magnetic ordering down to 1.8 K for \textit{R} = Tm and Yb \cite{pokharel2022highly,pokharel2021electronic,zhang2022electronic,guo2023triangular}. These materials exhibit exotic phenomena, such as quantum oscillations in YV$_{6}$Sn$_{6}$ \cite{pokharel2021electronic}, charge density waves in ScV$_{6}$Sn$_{6}$ \cite{zhang2022destabilization,arachchige2022charge}, and quantum critical behavior in YbV$_{6}$Sn$_{6}$ \cite{guo2023triangular}, underscoring the pivotal role of the \textit{R} sublattice in tuning both magnetic and electronic properties. Recent experimental studies on GdV$_{6}$Sn$_{6}$ have revealed ferromagnetic order at 5 K, along with a band structure suggestive of hosting nontrivial topological phases \cite{ishikawa2021gdv6sn6}. Additionally, ScV$_{6}$Sn$_{6}$ exhibits a charge-density-wave phase transition reminiscent of the CsV$_{3}$Sb$_{5}$ system \cite{arachchige2022charge}. Angle-resolved photoemission spectroscopy experiments on HoV$_{6}$Sn$_{6}$ and GdV$_{6}$Sn$_{6}$ have unveiled characteristic electronic features of a Kagome lattice, including Dirac cones, saddle points, and flat bands originating from the purely vanadium Kagome layer \cite{peng2021realizing}. Furthermore, the topologically nontrivial Dirac surface states in GdV$_{6}$Sn$_{6}$ have been manipulated to cross the Fermi energy \cite{hu2022tunable}, highlighting the potential of \textit{R}V$_{6}$Sn$_{6}$ family as an ideal platform for studying Kagome physics.
	
Motivated by the promising prospects offered by Vanadium-based Kagome lattices, we have embarked on synthesizing and characterizing \textit{R}V$_{6}$Sn$_{6}$ (\textit{R} = Tb, Dy, Ho, Er) compounds. Unlike Mn-based Kagome metals, these materials exhibit long-range magnetic order at significantly lower temperatures, below 4 K. Previous investigations \cite{lee2022anisotropic,zhang2022electronic} employing magnetization and heat capacity measurements have clearly indicated the presence of magnetic order for all the compounds proposed for this study. Our recent neutron diffraction measurements have revealed a reduction in the ordered moment compared to the anticipated free ion value for \textit{R}$^{3+}$ \cite{zhou2024ground}. In order to comprehend the ground state properties of these materials, we conducted nuclear heat capacity analysis and muon spin relaxation measurements. Our investigation reveals evidence of dynamical magnetism in the magnetically ordered ground state down to low temperatures, along with the magnetic ordering of a reduced magnetic moment.
	
\section{Experimental details}
Single crystals \textit{R}V$_{6}$Sn$_{6}$ (\textit{R}= Tb, Dy, Ho, Er) were prepared by the self-flux method as described previously \cite{zhou2024ground}. The phase purity and crystal structure were characterized using single-crystal X-ray diffraction measurements, and the data were refined using the Jana2006 program \cite{petvrivcek2014crystallographic}. The obtained structural parameters closely matched those previously reported for the same compound in Ref.\cite{zhou2024ground}, confirming the formation of hexagonal structure with the P6/mmm space group. The heat capacity measurements were performed down to 50 mK at the National Cheng Kung University (NCKU), Taiwan using a PPMS from Quantum Design. Zero-field (ZF) and longitudinal-field (LF) $\mu$SR measurements were conducted on the co-aligned single crystals (with a mass of approximately $\sim$200 mg) at the $\pi$M3 beamline at Paul Scherrer Institute Villigen, Switzerland using the low-background GPS instrument. The sample was mounted on a Cu holder with GE-varnish and covered with an aluminum foil of 10 $\mu$m thickness. It was then mounted in a He-gas continuous-flow cryostat refrigerator (CCR), with the c-axis parallel to the muon beam direction, to cover the temperature range between 1.5 and 200 K. 
		
\section{Experimental Results}

\subsection{Heat capacity measurements and analysis}

\begin{figure}[htbp]  \vspace{-0.2cm}
\begin{center} 
\includegraphics[scale=0.5]{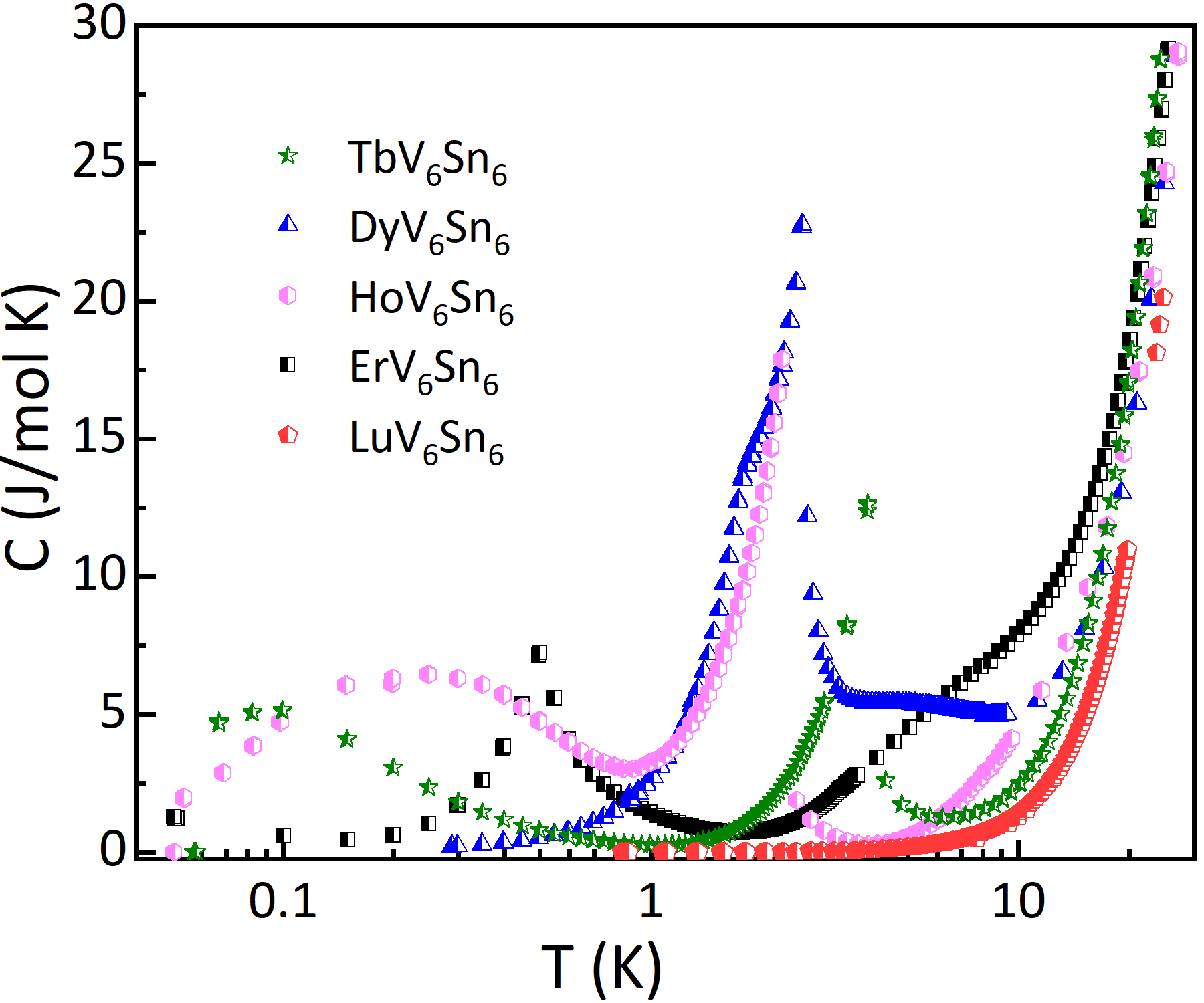}
\vspace{-5pt}
\caption{Temperature dependence of the total specific heat \textit{C}$_{p}$(\textit{T}) of \textit{R}V$_{6}$Sn$_{6}$ (\textit{R}= Tb, Dy, Ho, Er and Lu) compounds measured at zero field down to 50 mK.} 
\end{center} \vspace{-10pt}
\end{figure}  \noindent

We first present specific heat $C_{p}$(\textit{T}) data, indicating long-range magnetic ordering anomaly (see Fig. 1) at \textit{T$_{C}$} = 4.3 K, 3 K and 2.4 K for TbV$_{6}$Sn$_{6}$, DyV$_{6}$Sn$_{6}$, HoV$_{6}$Sn$_{6}$ and \textit{T$_{N}$} = 0.6 K for ErV$_{6}$Sn$_{6}$ respectively. This feature is in good agreement with magnetization and neutron diffraction results \cite{zhou2024ground}. Alongside the long-range magnetic transition, a pronounced upturn in the magnetically ordered state below 0.5 K is observed in HoV$_{6}$Sn$_{6}$, ErV$_{6}$Sn$_{6}$, and TbV$_{6}$Sn$_{6}$ systems. This characteristic feature is attributed to a nuclear Schottky anomaly, resulting from the energy level splitting of the \textit{R}$^{3+}$ nuclear spin induced by the hyperfine field originating from the electronic moment of the \textit{R}$^{3+}$ ions. Notably, the absence of magnetic order in (Lu/Y)V$_{6}$Sn$_{6}$ compared to TbV$_{6}$Sn$_{6}$ suggests that the magnetic order solely arises from the cooperative alignment of the 4$f$ electrons of Tb$^{3+}$ ions, indicating minimal influence from the nonmagnetic vanadium Kagome layers on magnetic behavior \cite{rosenberg2022uniaxial}. Consequently, it implies that there is no additional contribution to the hyperfine splitting, apart from that of the electron moment of the \textit{R}$^{3+}$ ions.\\
  
\begin{figure}[htbp]
\begin{center}     
\includegraphics[width=6cm,height=10cm]{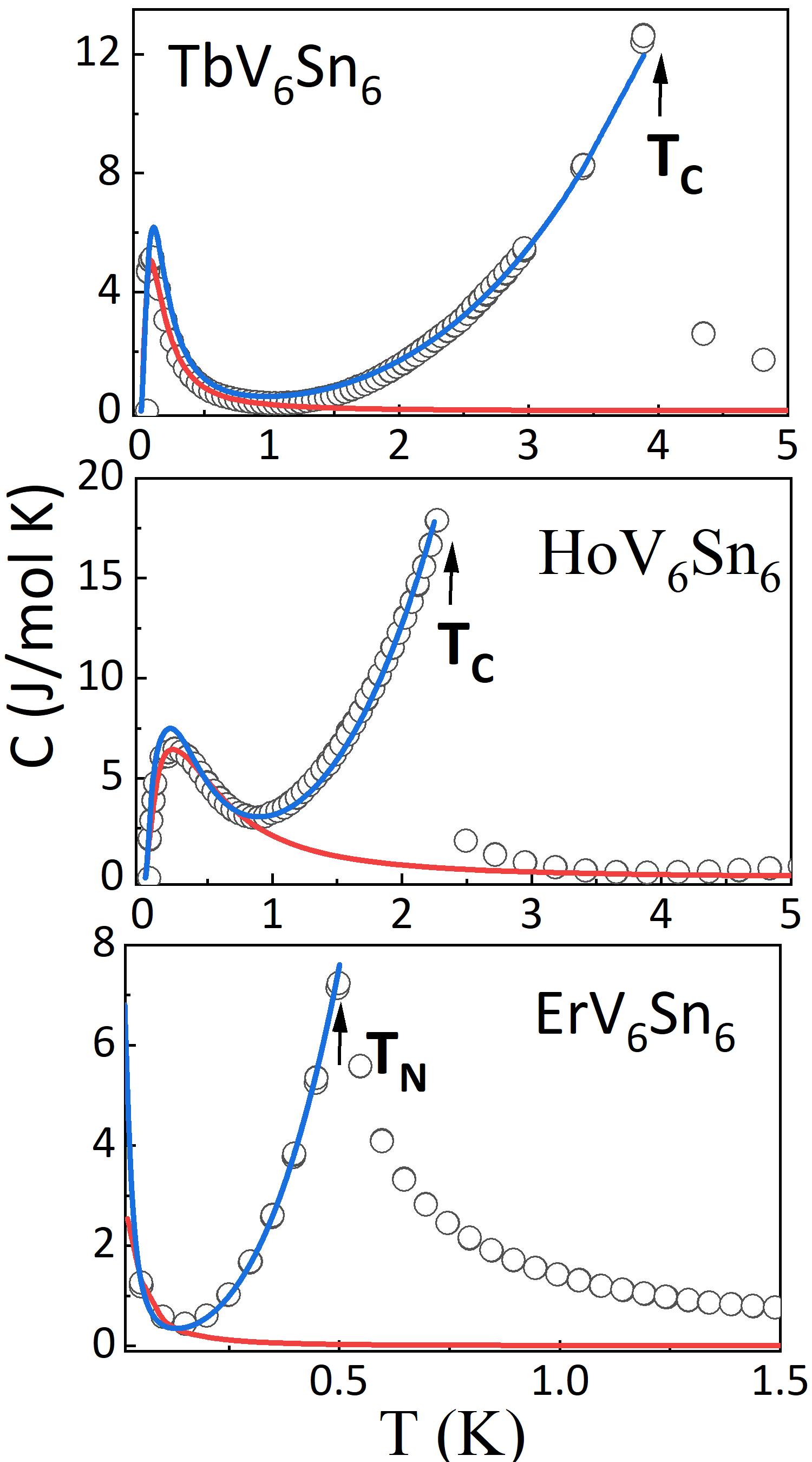}
\vspace{-5pt}
\caption{(Symbol) Specific heat \textit{C}$_{p}$ in TbV$_{6}$Sn$_{6}$, HoV$_{6}$Sn$_{6}$ and ErV$_{6}$Sn$_{6}$ systems. The curve below \textit{T}$_{C/N}$ is fitted with the sum of \textit{T}$^{3}$ and of a nuclear Schottky anomaly, the later being also computed with hyperfine model for the magnetically ordered system. The arrow label the long-range magnetic ordering temperature.} 
\end{center} 	 \vspace{-15pt}
\end{figure} 
\noindent

To elucidate the characteristics of the magnetically ordered state, the nuclear heat capacity was analyzed using the hyperfine model \cite{bloch1973hyperfine} 
\begin{eqnarray*}
C_{N} =\frac{R\sum\limits_{i=-I}^{I}\sum\limits_{j=-I}^{I}(W_{i}^{2}-W_{i}W_{j})e^{\frac{-W_{i}-W_{j}}{kT}}}{(kT)^{2}\sum\limits_{i=-I}^{I}\sum\limits_{j=-I}^{I}e^{\frac{-W_{i}-W_{j}} {kT}}} 
\end{eqnarray*}
\noindent

where \textit{R} is the universal gas constant, \textit{I} is the nuclear spin. The energy levels W$_{i}$ are given by, \textit{W}$_{i}$/\textit{k} = \textit{a}$^{\prime}$\textit{i} + \textit{P}[\textit{i}$^{2}$ - \textit{I}(\textit{I}+1)/3] where i = -\textit{I}, -\textit{I}+1,-\textit{I}+2,....,\textit{I}+2,\textit{I}-1,\textit{I}.
Here, \textit{a}$^{\prime}$ is a measure of the strength of hyperfine interactions and \textit{P} is the quadrupole coupling constant. For many rare-earth based systems, this method has been used to successfully estimate the nuclear contribution in thermal analysis \cite{bramwell2001spin,blote1969heat,nagata2001specific}. The obtained nuclear hyperfine contribution $C_{Nuc}$ is shown in Fig. 2 (red line). The best fit yields values of hyperfine coupling constant as \textit{a}$^{\prime}$= 0.26 K, 0.15 K, and 0.01 K for HoV$_{6}$Sn$_{6}$, TbV$_{6}$Sn$_{6}$, and ErV$_{6}$Sn$_{6}$, respectively. Notably, the obtained parameters for all compounds, except ErV$_{6}$Sn$_{6}$, agree well with previous reports \cite{blote1969heat,stevens1976nuclear,kumar2016hyperfine,krusius1969calorimetric,krusius1974hyperfine,neogy1979investigation,sheetal2022field}. The significant reduction in the a$^{\prime}$ for ErV$_{6}$Sn$_{6}$ is primarily attributed to the constrained temperature range, preventing full access to the nuclear heat capacity peak and thereby affecting the accuracy of the fitting results.\\
	 
An alternative estimate of the nuclear term was obtained using the magnetic hyperfine interactions in the magnetically ordered state (shown in Fig. 2 (blue line)). This is given by $\mathcal{H}$ = -m$_{n}$.H$_{hf}$, where the nuclear moment m$_{n}$ = $\mu_{n}$$g_{n}$I and the hyperfine field H$_{hf}$ = C$\mu$, C being the hyperfine proportionality constant and $\mu$ is the rare-earth moment. Let $\Delta$ be the common energy splitting between adjacent hyperfine levels and \textit{z} = $\Delta$/2k$_{B}$T. Considering the equidistant level scheme and the nuclear Schottky anomaly solely influenced by the magnetic hyperfine interaction, then the specific heat of the Schottky anomaly for one mole is described by \cite{bonville2010spin,stevens1976nuclear} 
	
\begin{eqnarray*}
C_{Nuc} = Rz^{2}\bigg[\frac{1}{sinh^{2}(z)} - \bigg( \frac{2I+1}{sinh[z(2I+1)]}\bigg)^{2}\bigg]
\end{eqnarray*}
		\\
where $\Delta$ = $\mu_{n}$/I C$\mu$, m$_{n}$ = 2$\mu_{N}$, I = 3/2 for Tb, 4.17$\mu_{N}$, 7/2 for Ho and 0.562$\mu_{N}$, 7/2 for Er. For rare-earth materials characterized by highly localized magnetism, the hyperfine constant remains independent of the compound. Thus, by utilizing this approach \cite{ryan2003166, mirebeau2005ordered,bloch1973hyperfine}, we deduce the magnetic moments as \textit{m}$_{eff}$ = 5.46(3)$\mu_{Ho}$, 9.05(4)$\mu_{Tb}$ and 1.90(5)$\mu_{Er}$. The magnetic moments for Tb and Ho are slightly closer to the neutron derived moment of 9.4(2)$\mu_{Tb}$ and 6.4(2)$\mu_{Ho}$ and comparatively lesser than the free ion value of \textit{R}$^{3+}$ ions. The estimated moment of 1.90(5)$\mu_{Er}$ is significantly smaller than both the neutron-derived moment of 6.1(3) $\mu_{Er}$ and the free ion value of the Er$^{3+}$ ion, which is 9.58 $\mu_{B}$/Er. This substantial reduction in moment may be attributed to the limited accessible temperature range. However, an alternative possibility is the presence of fluctuations persisting down to low temperatures, which could influence the magnetic ordering consistent with the reduced value obtained from neutron diffraction \cite{zhou2024ground}. Recent neutron studies suggest the existence of XY anisotropy in Er and Heisenberg-type ordering in Er, Ho, and Dy-based systems \cite{zhou2024ground}. The influence of fluctuations often leads to a reduction in the electronic magnetic moments that affect the Zeeman splitting of nuclear magnets \cite{mirebeau2005ordered,bonville2010spin}. The apparent decrease in the moment implies the presence of spin fluctuations within the magnetically ordered ground state. Spin fluctuations can be associated with highly anisotropic magnetization, where fluctuations in the transverse component of the moment likely reduce the overall magnetic moment of the system \cite{zhou2024ground}. It is important to note that while crystal field effects set the electronic moment at higher temperatures, their impact is evident in the reduced magnetic moment observed at low temperatures, alongside factors like spin fluctuations and anisotropy \cite{pokharel2022highly, pokharel2021electronic}. At present without inelastic neutron scattering, the exact information of the complex crystal field scheme extending to high temperatures cannot be precisely estimated.

\subsection{$\mu$SR measurements and analysis}
			
	\begin{figure*}[htbp]
				\begin{center} 
					\includegraphics[scale=0.55]{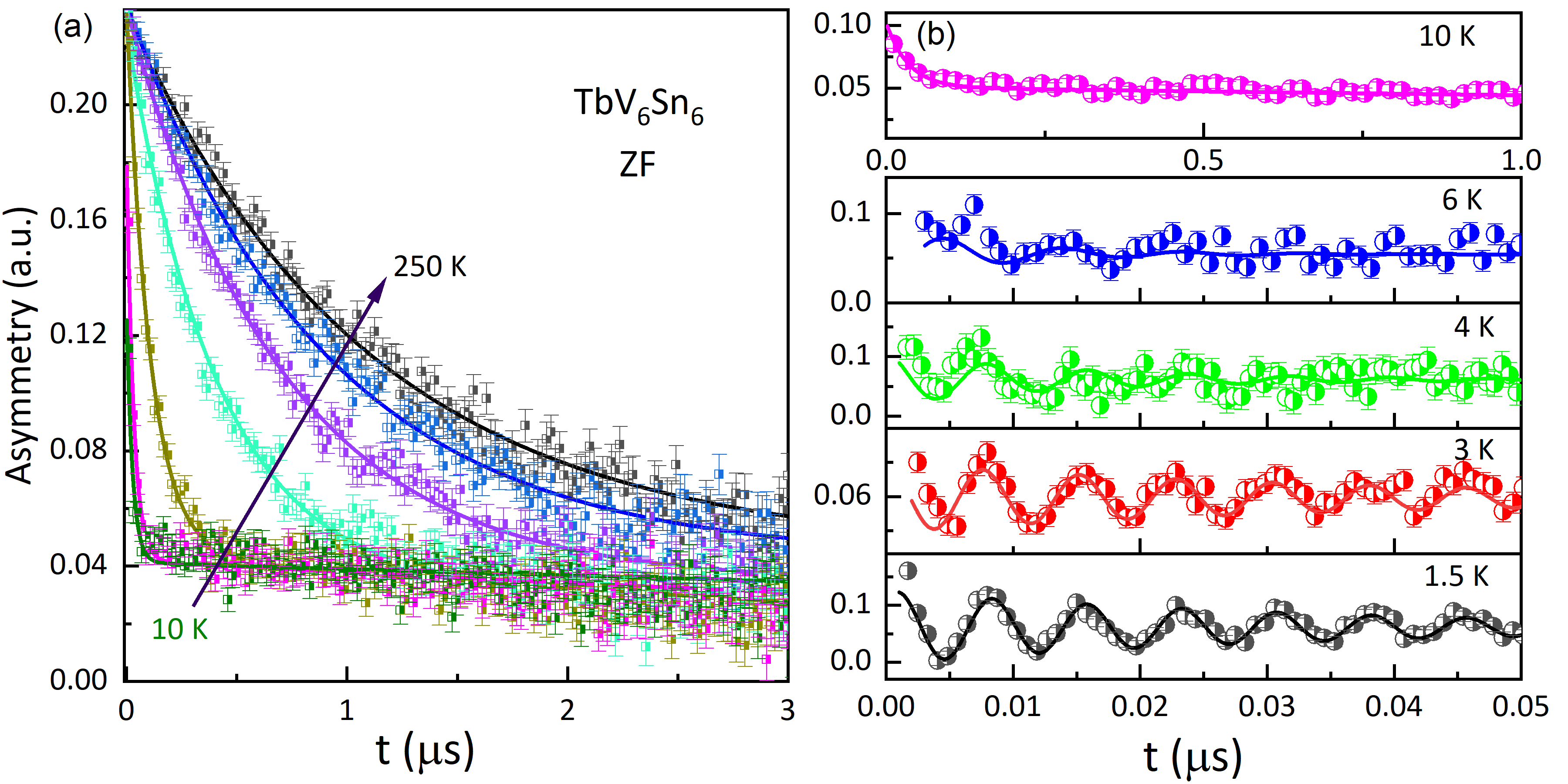}
					\includegraphics[scale=0.55]{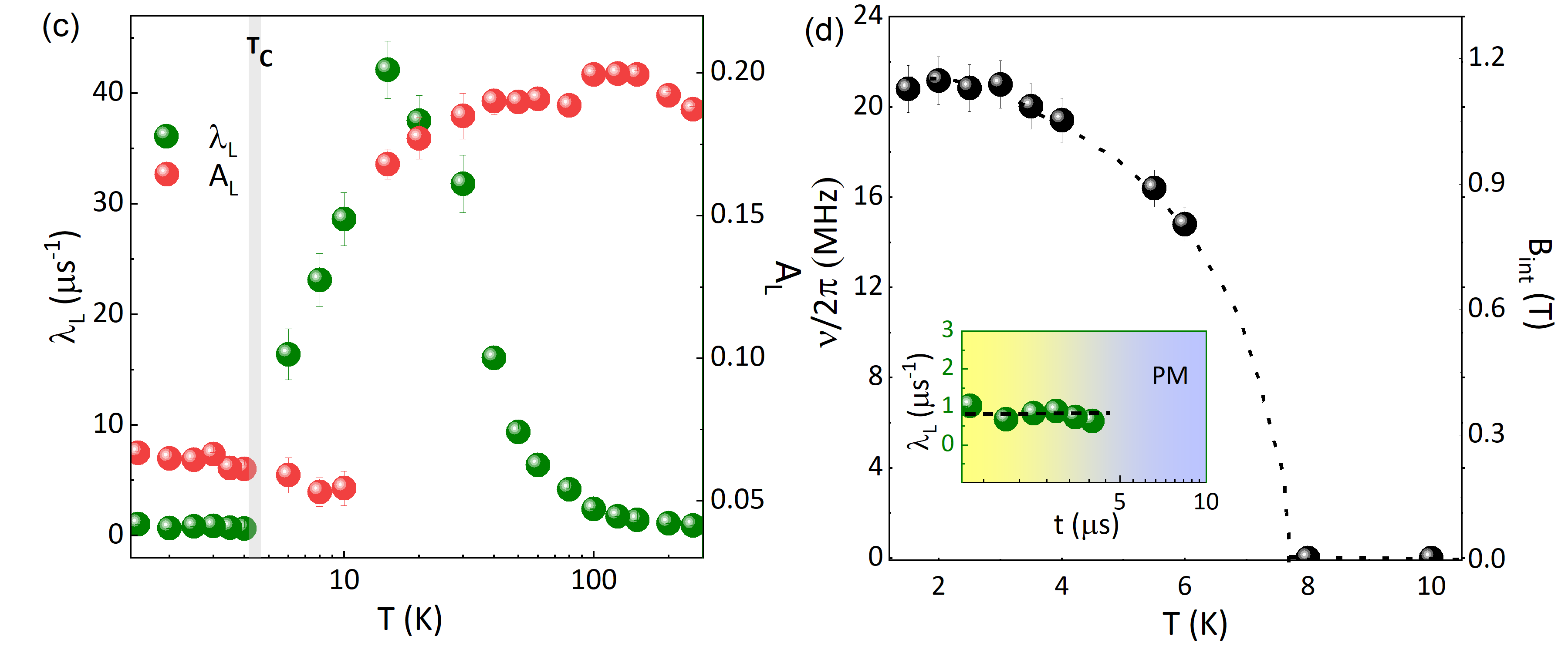}
					\caption{ (a and b) Shows ZF $\mu$SR spectra of TbV$_{6}$Sn$_{6}$ collected at various temperatures, showing clear oscillations as \textit{T}$\rightarrow$\textit{T}$_{C}$ (\textit{T}$_{C}$ taken from heat capacity). Solid lines represent the fit as described in the text. (c and d) Shows the temperature dependence of the fit parameters relaxation rate $\lambda_{L}$, asymmetry \textit{A}$_{L}$, frequency $\nu/2\pi$ and mean field at muon site B$_{int}$  extracted by fitting the zero field data collected between 1.5 and 250 K. The dotted line is a guide to the eye. The inset of Fig. 3d displays the temperature-independent longitudinal relaxation rate, $\lambda_L$, in the magnetically ordered regime, which remains finite at approximately 1 $\mu$s$^{-1}$.}
				\end{center} 	\vspace{-15pt}
	\end{figure*}
			
To gain further insight into the nature of the magnetic ground state and associated spin dynamics, we performed ZF and LF $\mu$SR measurements on DyV$_{6}$Sn$_{6}$ and TbV$_{6}$Sn$_{6}$ samples. The ZF spectra between 250 K and 1.5 K are shown in Figures 3a and 3b for TbV$_{6}$Sn$_{6}$. As the temperature decreases below 20 K, we observe a loss of initial asymmetry and the onset of spontaneous muon precession below \textit{T$_{C}$} = 4.3 K. Over longer timescales, the relaxation reflects spin dynamics associated with the longitudinal component of the internal magnetic field, which is parallel to the muon spin polarization. In contrast, on shorter timescales (\textit{t} $<$ 0.1 $\mu$s), the presence of damped oscillations signifies a distribution of quasi-static internal magnetic fields oriented perpendicular to the muon spin polarization. Therefore, the presence of well-defined muon oscillations clearly distinguishes the ferromagnetically ordered state from the paramagnetic state. Weak oscillations observed above \textit{T$_{C}$} indicate the presence of coherent local fields, suggesting the onset of magnetic correlations at temperatures higher than the nominal bulk ordering temperature. This implies that spin correlations develop prior to the establishment of full long-range order. The concomitant peak in longitudinal relaxation (Fig. 3c) reflects the slowing down of spin fluctuations in this regime.

The ZF spectra in the paramagnetic regime were fitted using a simple exponential function \textit{A(t)} = \textit{A}$_{L}$exp$^{(-\lambda_{L}t)}$ and a constant background term estimated at high temperature. Here, A$_{L}$ is the initial asymmetry and $\lambda_{L}$ represents the relaxation rate corresponding to the dynamic fields at muon sites. For \textit{T} $<$ \textit{T}$_{C}$, an additional term is added to the relaxation function for fitting the oscillation part, \textit{A(t)} = \textit{A}$_{L}$exp$^{-(\lambda_{L}t)}$ + \textit{A}$_{T}$exp$^{-(\lambda_{T}t)}$cos($\omega$t + $\phi$). Unlike TbMn$_{6}$Sn$_{6}$ \cite{mielke2022low}, where the muon experiences different magnetic environments corresponding to two magnetically inequivalent sites, the nature of the oscillations in the magnetically ordered state of TbV$_{6}$Sn$_{6}$ is well-described using a single precession frequency. The fit parameters obtained are shown in Figure 3c and 3d. A transition at \textit{T}$_{C}$ is evident from the temperature dependence of $\lambda_{L}$ and \textit{A}$_{L}$, where both parameters drop sharply upon entering the magnetically ordered state. Furthermore, the temperature dependence of the internal field (B$_{int}$), determined from the oscillation frequency in the muon asymmetry data (\textit{f}$_{\mu}$ = ($\gamma$$_\mu$/2$\pi$) B$_{int}$), is presented in Fig. 3d. B$_{int}$ is essentially zero at 8 K, becomes finite at 6 K, and reaches a saturation value of approximately 10 kOe below the ordering temperature T$_{C}$ = 4.3 K, comparable to that observed in TbMn$_6$Sn$_6$ \cite{mielke2022low}. We note an important feature in the temperature dependence of the longitudinal spin relaxation rate, $\lambda_{L}$, which captures the dynamics of magnetic fluctuations. In a fully ordered magnetic system, $\lambda_{L}$$\rightarrow$0 as $T$$\rightarrow$0, reflecting the approach of magnetization toward a static limit. In contrast, $\lambda_{L}$ in TbV$_6$Sn$_6$ remains nearly constant at 1 $\mu$s$^{-1}$ (as shown in Fig. 3d), demonstrating low-temperature magnetic fluctuations within the ordered state. Such fluctuations likely account for the reduced ordered moment inferred from hyperfine field analysis and previously reported neutron diffraction data \cite{zhou2024ground}.\\
	
	\begin{figure*}[htbp]
		\begin{center} 
			\includegraphics[scale=0.7]{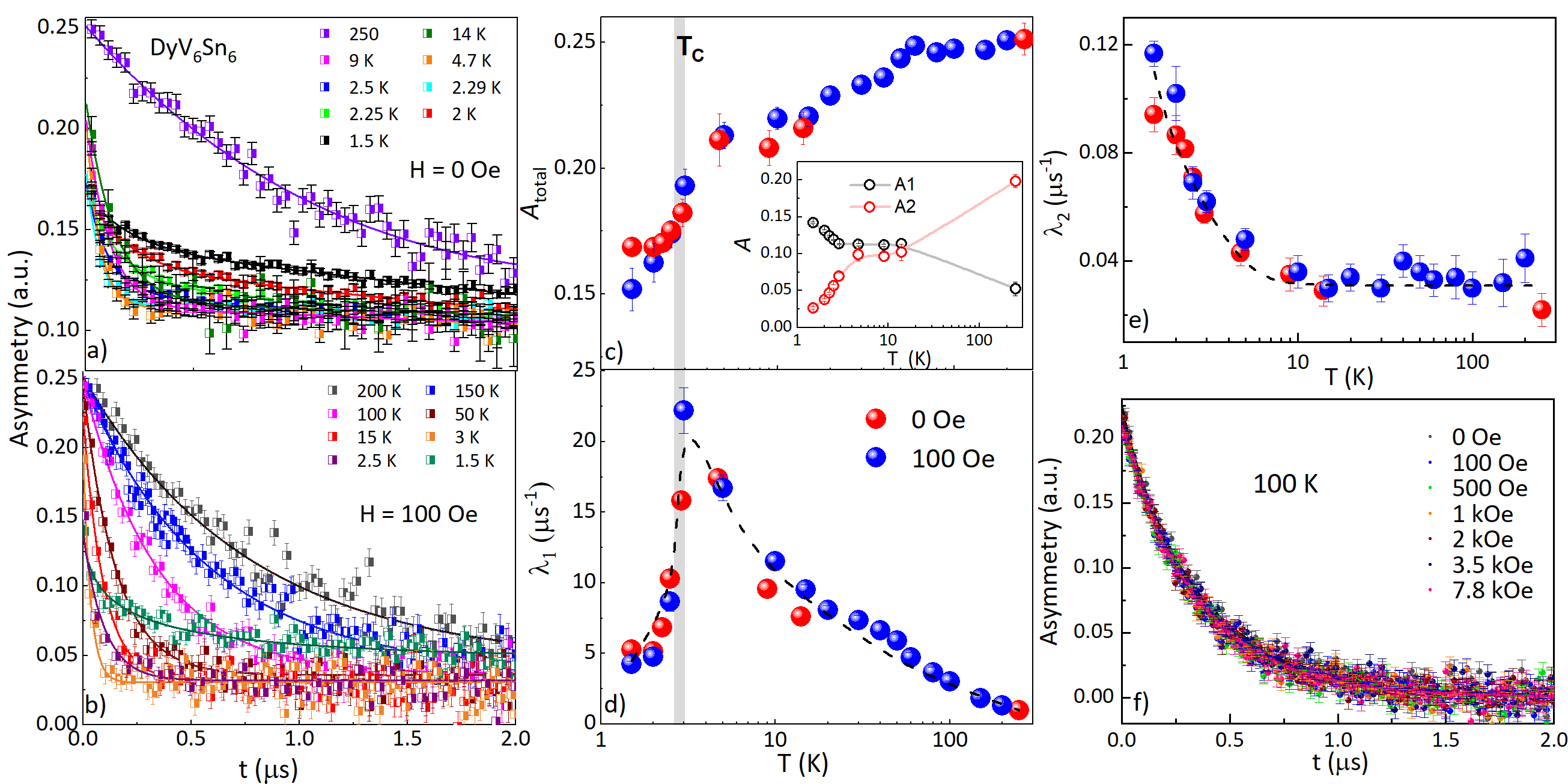}
			\caption{(a, b) ZF and 100 Oe $\mu$SR spectra of DyV$_{6}$Sn$_{6}$ at various temperatures. Solid lines are fitting results as explained in the main text. (c, d and e) Shows the temperature dependence of fit parameters relaxation rate $\lambda_1$, $\lambda_2$, total asymmetry and the inset of Fig. 4c shows temperature dependence of asymmetry paramters for ZF. The grey region marked the long-range ordering temperature from the heat capacity measurements. (4f) Shows the $\mu$SR spectra collected at various longitudinal fields ranging between 0 and 7.8 kOe at 100 K.}
		\end{center}  \vspace{-10pt}
	\end{figure*} 
	
	Next we will discuss the $\mu$SR results on DyV$_{6}$Sn$_6$. ZF spectra collected at several temperatures are shown in Fig. 4a, while data recorded in a small longitudinal field (100 Oe) and in a broader field range at higher temperature are displayed in Fig. 4b and Fig. 4f, respectively. Unlike TbV$_6$Sn$_6$, coherent oscillations associated with well-defined muon precession are not observed in DyV$_6$Sn$_6$ below the bulk ordering temperature $T_C$ $\approx$ 2.9 K. The lack of observable oscillations is usually caused by the presence of very large quasi-static local fields produced by large magnetic moments. These large fields depolarize the muon spin on a timescale shorter than the spectrometer time resolution, suppressing detectable precession even when long-range magnetic order is present \cite{lee2022anisotropic,zhang2022electronic,zhou2024ground}. However, neutron diffraction finds an ordered Dy moment of 6.6(2) $\mu_B$ \cite{zhou2024ground}; scaling from the internal field in TbV$_6$Sn$_6$ yields an estimated local field at the muon site of order $\sim$7 kOe for DyV$_6$Sn$_6$, which is actually smaller than that in TbV$_6$Sn$_6$. Given that the ordered magnetic moment in DyV$_6$Sn$_6$ is slightly canted from the crystallographic $c$-axis \cite{zhou2024ground}, this non-collinear magnetic order may generate complex and fluctuating local-field condition at the muon site, thus leading to the lost coherent oscillations.
	
	The ZF and low-LF spectra were modeled using a phenomenological two-exponential function $A(t) = A_1\,e^{-\lambda_1 t} \;+\; A_2\,e^{-\lambda_2 t} \;+\; A_{\mathrm{BG}}$, where $A_1$ and $A_2$ are the asymmetries of the fast and slow components, $\lambda_1$ and $\lambda_2$ their associated relaxation rates and $A_{\mathrm{BG}}$ a temperature-independent background. This two-component relaxation reflects two distinct muon stopping environments: a majority fraction sensing large quasi-static Dy fields and a minority fraction experiencing weaker, slowly fluctuating fields. This two-component form provides an excellent description of the data across the measured temperature range (see Fig. 4a, b) and yields internally consistent parameters for both ZF and 100 Oe datasets. The temperature dependence of the total sample asymmetry ($A_1+A_2$) evolves smoothly through $T_C$ (Fig. 4c). The dominant fast component $\lambda_1$ exhibits a pronounced peak near $T_C$ (Fig. 4d), a hallmark of critical slowing down as spin fluctuations slow on approaching the transition. Below $T_C$, $\lambda_1$ remains large but decreases gradually on cooling, consistent with the development of a quasi-static, spatially inhomogeneous internal-field distribution associated with Dy-moment ordering. By contrast, the slow component $\lambda_2$ is small, shows no sharp anomaly at $T_C$ and increases only weakly with decreasing temperature; this behavior is naturally interpreted as a weakly fluctuating or quasi-static background contribution, which may arise from muons stopping in regions of reduced local field. The persistence of a substantial low-temperature $\lambda_1$ $\approx$ 5 $\mu$s$^{-1}$ indicates that, even within the ordered phase, slow spin dynamics or a broad distribution of static fields remain important on the muon timescale. LF measurements up to 7.8 kOe were performed at 100 K (well above $T_C$) to probe the field dependence of the relaxation (Fig. 4f). No full recovery of the initial asymmetry was observed within this field range, indicating that the internal fields at 100 K remain predominantly dynamic and are not completely decoupled by these longitudinal fields \cite{devi2024muon}. This observation implies that higher fields (or lower temperatures) would be required to fully quench the dynamic contribution and to better resolve any remaining static fraction.

\section{Discussions}

The \textit{R}V$_6$Sn$_6$ (\textit{R} = Tb, Dy, Ho, Er) compounds investigated here exhibit a coexistence of long-range magnetic order and low-energy spin dynamics below 4 K. Heat-capacity data confirm well-defined magnetic transitions, while hyperfine and $\mu$SR measurements show that slow fluctuations persist deep inside the ordered state. A notable feature is the reduced ordered moment inferred from the nuclear hyperfine analysis, most prominently in ErV$_6$Sn$_6$, where the value is significantly smaller than that obtained from neutron diffraction \cite{zhou2024ground}. This reduction is consistent with an incomplete resolution of the nuclear Schottky anomaly together with strong CEF anisotropy and residual slow dynamics, both of which can weaken the static hyperfine field at millikelvin temperatures. Classical analyses of rare-earth systems further establish that Kramers ions like Er$^{3+}$ naturally produce sizeable nuclear contributions under partial moment freezing \cite{blote1969heat,bleaney1963hyperfine, dasgupta2006crystal}. These parallels support a predominantly nuclear origin for the low-temperature heat-capacity upturn observed in ErV$_6$Sn$_6$.\\

The $\mu$SR measurements provide complementary, time-resolved insight into the magnetic dynamics. In TbV$_6$Sn$_6$, coherent oscillations appear below $T_C$, indicating a well-defined quasistatic internal field at the muon site. The longitudinal relaxation rate $\lambda_L$ remains finite and weakly temperature-dependent as $T \rightarrow 0$, demonstrating that slow spin fluctuations persist within the ordered state. This behavior is consistent with a reduced ordered moment and suggests that anisotropic exchange or transverse CEF components prevent complete static ordering of the Tb moments. In DyV$_6$Sn$_6$, the $\mu$SR response is qualitatively different. The absence of coherent oscillations together with extremely rapid early-time depolarization points to very large, spatially inhomogeneous internal fields generated by the Dy$^{3+}$ moments. Although no oscillations are resolved, the fast-relaxing component remains sizable below $T_C$, showing that slow fluctuations or field inhomogeneities persist within the ordered state. The contrast between Tb and Dy highlights the strong dependence of the local magnetic environment on the rare-earth ion, reflecting variations in moment magnitude, single-ion anisotropy, and \textit{R}--\textit{R} exchange pathways. Finally, we note that the present $\mu$SR data for DyV$_6$Sn$_6$ extend down to 1.5 K. The finite relaxation rates observed well below $T_C$ indicate that slow spin dynamics remain active within the ordered phase. Measurements at lower temperatures, together with systematic longitudinal-field decoupling studies, would clarify whether these fluctuations persist to the lowest energies or gradually freeze out, and would help disentangle intrinsic slow dynamics from quasi-static field inhomogeneity.

It is instructive to compare these results with those of Mn-based Kagome compounds \textit{R}Mn$_6$Sn$_6$. In the Mn systems, the magnetic energy scale is dominated by the 3$d$ Mn sublattice, resulting in high ordering temperatures and strong 3$d$--4$f$ coupling \cite{venturini1991magnetic,venturini1996incommensurate,malaman1999magnetic,wang2023real,sales2019electronic,ghimire2020competing}. By contrast, the V-based compounds studied here host magnetism solely on the rare-earth sublattice, while the V Kagome layers remain nonmagnetic \cite{lee2022anisotropic,zhou2024ground}. This leads to reduced ordering temperatures and enhanced sensitivity to CEF anisotropy, layered structure, and weak interlayer exchange. These factors enable slow spin dynamics to coexist with long-range order, particularly in DyV$_6$Sn$_6$ where strong anisotropy produces broad internal-field distributions. Overall, the \textit{R}V$_6$Sn$_6$ series represents a clean platform for studying anisotropy-driven magnetism and low-energy fluctuations in rare-earth Kagome systems. The coexistence of magnetic order with persistent slow dynamics revealed by heat-capacity and $\mu$SR measurements underscores the importance of CEF effects, exchange anisotropy, and moment-dependent variability across the series. Further single crystal inelastic neutron scattering and dilution-temperature $\mu$SR investigations will be essential to map the CEF schemes and clarify the microscopic origin of these unconventional low-temperature dynamics.
    	     
\section{Conclusions}
In summary, combined heat capacity and $\mu$SR measurements establish the presence of long-range magnetic order across all \textit{R}V$_{6}$Sn$_{6}$ (\textit{R} = Tb, Dy, Ho, Er) compounds below 4 K. Despite the development of static order, persistent low-temperature spin fluctuations remain evident, indicating coexistence of ordered and dynamic magnetism. The reduced magnetic moments derived from hyperfine analysis further support this dynamic character within the ordered phase. These results demonstrate that the magnetism in V-based Kagome metals is governed primarily by the anisotropic 4$f$ moments of the rare-earth ions, leading to diverse ground states shaped by competing interactions and local anisotropy. Compared with Mn-based Kagome counterparts, the \textit{R}V$_{6}$Sn$_{6}$ series reveals a distinct magnetic landscape emerging from the weakly coupled rare-earth sublattice and the nonmagnetic V Kagome layers, offering a clean platform to explore anisotropy-driven spin dynamics in geometrically frustrated lattices.
		
\section*{Acknowledgment}	
	
This work is based on the $\mu$SR experiments performed at the GPS muon instrument (S$\mu$S, PSI, Villigen). The single-crystal growth was performed at JCNS-MLZ, Garching. The heat capacity was measured with a PPMS device and a dilution refrigerator system at NCKU. We would like to acknowledge P. Bonville for fruitful discussions, and C. Baines, S. Hammouda, P. Pal, Y.H. Tung, P.C. Chang, J. Ramon for their assistances in the experiments. S.D. acknowledges the postdoctoral funding from the European Union's Horizon 2020 research and innovation programme under the Marie Skłodowska-Curie grant agreement No 101034266. Y.Z. acknowledges the scholarship funding from the Chinese Scholarship Council. 

\bibliography{Ref.bib}

@article{bleaney1963hyperfine,
	title={Hyperfine Interactions in Rare-Earth Metals},
	author={Bleaney, B},
	journal={Journal of Applied Physics},
	volume={34},
	number={4},
	pages={1024--1031},
	year={1963},
	publisher={American Institute of Physics}
}

@article{dasgupta2006crystal,
	title={Crystal field effect and geometric frustration in Er2Ti2O7—an XY antiferromagnetic pyrochlore},
	author={Dasgupta, Papri and Jana, Yatramohan and Ghosh, Debjani},
	journal={Solid state communications},
	volume={139},
	number={8},
	pages={424--429},
	year={2006},
	publisher={Elsevier}
}

@article{sales2019electronic,
	title={},
	author={Sales, Brian C and Yan, Jiaqiang and Meier, William R and Christianson, Andrew D and Okamoto, Satoshi and McGuire, Michael A},
	journal={Physical Review Materials},
	volume={3},
	number={11},
	pages={114203},
	year={2019},
	publisher={APS}
}

@article{ghimire2020competing,
	title={},
	author={Ghimire, Nirmal J and Dally, Rebecca L and Poudel, L and Jones, DC and Michel, D and Magar, N Thapa and Bleuel, M and McGuire, Michael A and Jiang, JS and Mitchell, JF and others},
	journal={Science Advances},
	volume={6},
	number={51},
	pages={eabe2680},
	year={2020},
	publisher={American Association for the Advancement of Science}
}

@article{zhou2024ground,
	title={},
	author={Zhou, Yishui and Lee, Min-Kai and Hammouda, Sabreen and Devi, Sheetal and Yano, Shin-Ichiro and Sibille, Romain and Zaharko, Oksana and Schmidt, Wolfgang and Schmalzl, Karin and Beauvois, Ketty and others},
	journal={Physical Review Research},
	volume={6},
	number={4},
	pages={043291},
	year={2024},
	publisher={APS}
}

@article{devi2024muon,
	title={},
	author={Devi, Sheetal and Biswas, Pabitra K and Yokoyama, K and Adroja, DT and Yadav, CS},
	journal={J. Phys. Condens. Matter},
	volume={36},
	number={34},
	pages={345802},
	year={2024},
	publisher={IOP Publishing}
}

@article{arachchige2022charge,
	title={},
	author={Arachchige, Hasitha W Suriya and Meier, William R and Marshall, Madalynn and Matsuoka, Takahiro and Xue, Rui and McGuire, Michael A and Hermann, Raphael P and Cao, Huibo and Mandrus, David},
	journal={Phys. Rev. Lett.},
	volume={129},
	number={21},
	pages={216402},
	year={2022},
	publisher={APS}
}

@article{zhang2022destabilization,
	title={},
	author={Zhang, Xiaoxiao and Hou, Jun and Xia, Wei and Xu, Zhian and Yang, Pengtao and Wang, Anqi and Liu, Ziyi and Shen, Jie and Zhang, Hua and Dong, Xiaoli and others},
	journal={Materials},
	volume={15},
	number={20},
	pages={7372},
	year={2022},
	publisher={MDPI}
}

@article{guo2023triangular,
	title={},
	author={Guo, Kaizhen and Ye, Junyao and Guan, Shuyue and Jia, Shuang},
	journal={Phys. Rev. B},
	volume={107},
	number={20},
	pages={205151},
	year={2023},
	publisher={APS}
}

@article{zhang2022electronic,
	title={},
	author={Zhang, Xiaoxiao and Liu, Ziyi and Cui, Qi and Guo, Qi and Wang, Ningning and Shi, Lifen and Zhang, Hua and Wang, Weihua and Dong, Xiaoli and Sun, Jianping and others},
	journal={Phys. Rev. Mater},
	volume={6},
	number={10},
	pages={105001},
	year={2022},
	publisher={APS}
}

@article{pokharel2021electronic,
	title={},
	author={Pokharel, Ganesh and Teicher, Samuel ML and Ortiz, Brenden R and Sarte, Paul M and Wu, Guang and Peng, Shuting and He, Junfeng and Seshadri, Ram and Wilson, Stephen D},
	journal={Phys. Rev. B},
	volume={104},
	number={23},
	pages={235139},
	year={2021},
	publisher={APS}
}

@article{pokharel2022highly,
	title={},
	author={Pokharel, Ganesh and Ortiz, Brenden and Chamorro, Juan and Sarte, Paul and Kautzsch, Linus and Wu, Guang and Ruff, Jacob and Wilson, Stephen D},
	journal={Phys. Rev. Mater},
	volume={6},
	number={10},
	pages={104202},
	year={2022},
	publisher={APS}
}

@article{xu2022topological,
	title={},
	author={Xu, Xitong and Yin, Jia-Xin and Ma, Wenlong and Tien, Hung-Ju and Qiang, Xiao-Bin and Reddy, PV Sreenivasa and Zhou, Huibin and Shen, Jie and Lu, Hai-Zhou and Chang, Tay-Rong and others},
	journal={Nat. commun.},
	volume={13},
	number={1},
	pages={1197},
	year={2022},
	publisher={Nature Publishing Group UK London}
}

@article{wang2022magnetotransport,
	title={},
	author={Wang, Bin and Yi, Enkui and Li, Leyi and Qin, Jianwei and Hu, Bing-Feng and Shen, Bing and Wang, Meng},
	journal={Phys. Rev. B},
	volume={106},
	number={12},
	pages={125107},
	year={2022},
	publisher={APS}
}

@article{ma2021rare,
	title={},
	author={Ma, Wenlong and Xu, Xitong and Yin, Jia-Xin and Yang, Hui and Zhou, Huibin and Cheng, Zi-Jia and Huang, Yuqing and Qu, Zhe and Wang, Fa and Hasan, M Zahid and others},
	journal={Phys. Rev. Lett.},
	volume={126},
	number={24},
	pages={246602},
	year={2021},
	publisher={APS}
}

@article{hu2023optical,
	title={},
	author={Hu, Tianchen and Pi, Hanqi and Xu, Shuxiang and Yue, Li and Wu, Qiong and Liu, Qiaomei and Zhang, Sijie and Li, Rongsheng and Zhou, Xinyu and Yuan, Jiayu and others},
	journal={Phys. Rev. B},
	volume={107},
	number={16},
	pages={165119},
	year={2023},
	publisher={APS}
}

@article{wang2020experimental,
	title={},
	author={Wang, Pengdong and Wang, Yihao and Zhang, Bo and Li, Yuliang and Wang, Sheng and Wu, Yunbo and Zhu, Hongen and Liu, Yi and Zhang, Guobin and Liu, Dayong and others},
	journal={Chinese Phys. Lett.},
	volume={37},
	number={8},
	pages={087102},
	year={2020},
	publisher={IOP Publishing}
}

@article{nagata2001specific,
	title={},
	author={Nagata, Shoichi and Sasaki, Hiromi and Suzuki, Kurando and Kiuchi, Junji and Wada, Nobuo},
	journal={J. Phys. Chem. Solids},
	volume={62},
	number={6},
	pages={1123--1130},
	year={2001},
	publisher={Elsevier}
}

@article{yin2020quantum,
	title={},
	author={Yin, Jia-Xin and Ma, Wenlong and Cochran, Tyler A and Xu, Xitong and Zhang, Songtian S and Tien, Hung-Ju and Shumiya, Nana and Cheng, Guangming and Jiang, Kun and Lian, Biao and others},
	journal={Nature},
	volume={583},
	number={7817},
	pages={533--536},
	year={2020},
	publisher={Nature Publishing Group UK London}
}

@article{blote1969heat,
	title={},
	author={Bl{\"o}te, HWJ and Wielinga, RF and Huiskamp, WJ},
	journal={Physica},
	volume={43},
	number={4},
	pages={549--568},
	year={1969},
	publisher={Elsevier}
}

@article{bramwell2001spin,
	title={},
	author={Bramwell, ST and Harris, MJ and Den Hertog, BC and Gingras, MJP and Gardner, JS and McMorrow, DF and Wildes, AR and Cornelius, AL and Champion, JDM and Melko, RG and others},
	journal={Phys. Rev. Lett.},
	volume={87},
	number={4},
	pages={047205},
	year={2001},
	publisher={APS}
}

@article{malaman1999magnetic,
	title={},
	author={Malaman, B and Venturini, G and Welter, R and Sanchez, JP and Vulliet, P and Ressouche, E},
	journal={Journal of magnetism and magnetic materials},
	volume={202},
	number={2-3},
	pages={519--534},
	year={1999},
	publisher={Elsevier}
}

@article{wang2023real,
	title={},
	author={Wang, Zhan and Xu, Jiawang and Li, Zhuolin and Xu, Tiankuo and Li, Jianqi and Zhao, Tongyun and Cai, Jianwang and Zhang, Ying and Shen, Baogen},
	journal={Applied Physics Letters},
	volume={122},
	number={11},
	year={2023},
	publisher={AIP Publishing}
}

@article{venturini1996incommensurate,
	title={},
	author={Venturini, G and Fruchart, D and Malaman, B},
	journal={Journal of alloys and compounds},
	volume={236},
	number={1-2},
	pages={102--110},
	year={1996},
	publisher={Elsevier}
}

@article{venturini1991magnetic,
	title={},
	author={Venturini, G and El Idrissi, B Chafik and Malaman, B},
	journal={Journal of magnetism and magnetic materials},
	volume={94},
	number={1-2},
	pages={35--42},
	year={1991},
	publisher={Elsevier}
}

@article{krusius1974hyperfine,
	title={},
	author={Krusius, M and Pickett, GR and Veuro, MC},
	journal={Solid State Commun.},
	volume={14},
	number={2},
	pages={191--193},
	year={1974},
	publisher={Elsevier}
}

@article{ryan2003166,
	title={},
	author={Ryan, DH and Cadogan, JM and Gagnon, R},
	journal={Phys. Rev. B},
	volume={68},
	number={1},
	pages={014413},
	year={2003},
	publisher={APS}
}

@article{neogy1979investigation,
	title={},
	author={Neogy, D and Chatterji, A},
	journal={J. Phys. Chem. Solids},
	volume={40},
	number={12},
	pages={1045--1049},
	year={1979},
	publisher={Elsevier}
}

@article{rosenberg2022uniaxial,
	title={},
	author={Rosenberg, Elliott and DeStefano, Jonathan M and Guo, Yucheng and Oh, Ji Seop and Hashimoto, Makoto and Lu, Donghui and Birgeneau, Robert J and Lee, Yongbin and Ke, Liqin and Yi, Ming and others},
	journal={Phys. Rev. B},
	volume={106},
	number={11},
	pages={115139},
	year={2022},
	publisher={APS}
}

@article{krusius1969calorimetric,
	title={},
	author={Krusius, M and Anderson, AC and Holmstr{\"o}m, B},
	journal={Phys. Rev.},
	volume={177},
	number={2},
	pages={910},
	year={1969},
	publisher={APS}
}

@article{bloch1973hyperfine,
	title={},
	author={Bloch, D and Voiron, J and Berton, A and Chaussy, J},
	journal={Solid State Commun.},
	volume={12},
	number={7},
	pages={685--687},
	year={1973},
	publisher={Elsevier}
}

@article{mielke2022low,
	title={},
	author={Mielke III, C and Ma, WL and Pomjakushin, V and Zaharko, O and Sturniolo, S and Liu, X and Ukleev, V and White, JS and Yin, J-X and Tsirkin, SS and others},
	journal={Commun. Phys},
	volume={5},
	number={1},
	pages={107},
	year={2022},
	publisher={Nature Publishing Group UK London}
}

@article{petvrivcek2014crystallographic,
	title={},
	author={Pet{\v{r}}{\'\i}{\v{c}}ek, V{\'a}clav and Du{\v{s}}ek, Michal and Palatinus, Luk{\'a}{\v{s}}},
	journal={Z Kristallogr Cryst Mater},
	volume={229},
	number={5},
	pages={345--352},
	year={2014},
	publisher={De Gruyter Oldenbourg}
}

@article{lee2022anisotropic,
	title={},
	author={Lee, Jeonghun and Mun, Eundeok},
	journal={Phys. Rev. Mater},
	volume={6},
	number={8},
	pages={083401},
	year={2022},
	publisher={APS}
}

@article{kumar2016hyperfine,
	title={},
	author={Kumar, CMN and Xiao, Y and Nair, HS and Voigt, J and Schmitz, Berthold and Chatterji, T and Jalarvo, NH and Br{\"u}ckel, Th},
	journal={J. Phys. Condens. Matter},
	volume={28},
	number={47},
	pages={476001},
	year={2016},
	publisher={IOP Publishing}
}

@inproceedings{bonville2010spin,
	title={},
	author={Bonville, P},
	booktitle={J Phys Conf Ser},
	volume={217},
	number={1},
	pages={012119},
	year={2010},
	organization={IOP Publishing}
}

@article{stevens1976nuclear,
	title={},
	author={Stevens, John Gehret and Dunlap, Bobby David},
	journal={J Phys Chem Ref Data},
	volume={5},
	number={4},
	pages={1093--1122},
	year={1976},
	publisher={American Institute of Physics for the National Institute of Standards and~…}
}

@article{mirebeau2005ordered,
	title={},
	author={Mirebeau, I and Apetrei, A and Rodr{\'\i}guez-Carvajal, J and Bonville, P and Forget, A and Colson, D and Glazkov, V and Sanchez, JP and Isnard, O and Suard, E},
	journal={Phys. Rev. Lett.},
	volume={94},
	number={24},
	pages={246402},
	year={2005},
	publisher={APS}
}

@article{hu2022tunable,
	title={},
	author={Hu, Yong and Wu, Xianxin and Yang, Yongqi and Gao, Shunye and Plumb, Nicholas C and Schnyder, Andreas P and Xie, Weiwei and Ma, Junzhang and Shi, Ming},
	journal={Sci. Adv},
	volume={8},
	number={38},
	pages={eadd2024},
	year={2022},
	publisher={American Association for the Advancement of Science}
}

@article{peng2021realizing,
	title={},
	author={Peng, Shuting and Han, Yulei and Pokharel, Ganesh and Shen, Jianchang and Li, Zeyu and Hashimoto, Makoto and Lu, Donghui and Ortiz, Brenden R and Luo, Yang and Li, Houchen and others},
	journal={Phys. Rev. Lett.},
	volume={127},
	number={26},
	pages={266401},
	year={2021},
	publisher={APS}
}

@article{ishikawa2021gdv6sn6,
	title={},
	author={Ishikawa, Hajime and Yajima, Takeshi and Kawamura, Mitsuaki and Mitamura, Hiroyuki and Kindo, Koichi},
	journal={J. Phys. Soc. Japan},
	volume={90},
	number={12},
	pages={124704},
	year={2021},
	publisher={The Physical Society of Japan}
}

@article{sheetal2022field,
	title={},
	author={Sheetal and Elghandour, A and Klingeler, R and Yadav, CS and others},
	journal={J. Phys. Condens. Matter},
	volume={34},
	number={24},
	pages={245801},
	year={2022},
	publisher={IOP Publishing}
}

\end{document}